\documentclass[preprint2]{aastex}
\usepackage{emulateapj5}
\usepackage{onecolfloat5}
\usepackage{epsfig}

\begin{document}


\def\egr{3EG~J1013--5915}
\def\psr{PSR~J1016--5857}
\def\pwn{G284.0--1.8}
\def\snr{G284.3--1.8}

\slugcomment{To appear in ApJ}
\shorttitle{CHANDRA DETECTION OF THE PWN AROUND PSR J1016--5857}
\shortauthors{CAMILO ET AL}

\twocolumn[
\title{{\em Chandra\/} Detection of a Synchrotron Nebula Around the
Vela-like Pulsar J1016--5857}

\author{F.~Camilo,\altaffilmark{1}
  B.~M.~Gaensler,\altaffilmark{2}
  E.~V.~Gotthelf,\altaffilmark{1}
  J.~P.~Halpern,\altaffilmark{1}
  and R.~N.~Manchester\altaffilmark{3}}
\altaffiltext{1}{Columbia Astrophysics Laboratory, Columbia University,
  550 West 120th Street, New York, NY~10027}
\altaffiltext{2}{Harvard-Smithsonian Center for Astrophysics, 60 Garden
  Street, Cambridge, MA~02138}
\altaffiltext{3}{Australia Telescope National Facility, CSIRO,
  P.O.~Box~76, Epping, NSW~1710, Australia}

\begin{abstract}
We report on a 19\,ks observation of the pulsar J1016--5857 with
the {\em Chandra X-ray Observatory}.  This ``Vela-like'' pulsar has
rotation period 107\,ms, characteristic age 21\,kyr, and spin-down power
$2.6\times10^{36}$\,ergs\,s$^{-1}$.  A relatively bright centrally peaked
source of radius $\approx 25''$ around the radio pulsar position has a
spectrum that is well fitted by an absorbed power law with photon index
$1.32\pm0.25$.  We regard this as a newly identified pulsar wind nebula
that we designate PWN~\pwn.  We do not detect the pulsar either as a
point X-ray source or as a pulsed source in a 55\,ks observation with
the {\em Rossi X-ray Timing Explorer\/} ({\em RXTE\/}).  The isotropic
PWN luminosity is $3\times10^{32}$\,ergs\,s$^{-1}$ in the 2--10\,keV
range, for a distance of 3\,kpc that is consistent with the measured
neutral hydrogen column density.  The unpulsed flux from the pulsar is
less than 30\% of the measured PWN flux.  The brightest component of
the PWN, near the pulsar, shows extended emission of size $\sim 2''$
that {\em may\/} indicate, by analogy with other young pulsars, the wind
termination shock.  In the {\em Chandra\/} image we also detect a very
faint extended structure $\approx 1'\times2'$ in size that is highly
asymmetric about the pulsar position.  This structure is a good match in
width and position angle to the tip of a ``finger'' of radio emission
that appears to connect to the nearby supernova remnant G284.3--1.8,
but we cannot characterize it further with the available data.
There is a variable X-ray point source only $1\farcm5$ from the pulsar,
which we identify using optical spectroscopy as an accreting binary.
\end{abstract}

]

\section{Introduction}\label{sec:intro} 

Magnetospherically active neutron stars with spin parameters similar
to those of the Vela pulsar have proven very useful to study a variety
of high-energy processes, such as thermal emission from the surface of
young neutron stars, pulsed emission at X-ray energies that can extend to
$\gamma$-ray energies and potentially power EGRET sources, interaction of
the relativistic pulsar wind with the immediate environment as observed
through a synchrotron nebula, and interaction of such a pulsar wind nebula
(PWN) or the pulsar with any existing supernova remnant (SNR).

\psr\ is one of about 25 ``Vela-like'' pulsars now known, typically
having spin periods $P \sim 0.1$\,s, characteristic ages $10^4\,{\rm yr}
\la \tau_c = P/2\dot P \la 10^5$\,yr, and spin-down luminosities $\dot E
= 4 \pi^2 I \dot P/P^3 \ga 10^{36}$\,ergs\,s$^{-1}$, where the moment of
inertia $I \equiv 10^{45}$\,g\,cm$^2$.  Like half of all Vela-like pulsars
known, it was discovered in the Parkes multibeam pulsar survey of the
Galactic plane \citep[e.g.,][]{mlc+01}.  Because pulsars discovered in
the multibeam survey are generally farther away than previously known
pulsars, they are expected to be fainter and more absorbed sources
of X-rays, and for the most part they have yet to be studied with
X-ray observatories.  \psr\ may be particularly interesting because
it appeared to be coincident with an {\em Einstein Observatory} X-ray
source, is positionally coincident with the very tip of a ``finger''
of radio emission apparently originating from the SNR~\snr, and is
also coincident with an unidentified EGRET source, \egr, that has
characteristics suggesting a possible pulsar origin \citep{cbm+01}.

For these reasons we obtained short observations with {\em Chandra\/}
and {\em RXTE\/} in order to investigate further the nature of the
environment of \psr.

\section{Observations, Analysis and Results}\label{sec:obs}

\subsection{{\em Chandra\/} Observation}\label{sec:cxo}

\psr\ was observed with {\em Chandra}'s Advanced CCD Imaging Spectrometer
(ACIS) on 2003 May 25 (MJD 52784).  The pulsar position obtained from
radio timing observations \citep{cbm+01} was placed on the aim-point
of the ACIS S3 CCD.  The data were collected in {\sc timed} exposure,
{\sc vfaint} mode, and the original event files were reprocessed with
up-to-date calibration files.  All {\em Chandra} data processing used
the latest {\sc ciao} tools.  There were no periods of abnormally high
background, and the effective integration time was 18,615\,s.

\subsection{{\em RXTE\/} Observation}\label{sec:rxte}

A 64,450\,s observation targeting \psr\ was obtained by {\em RXTE\/} on
2004 March 6 (MJD 53070).  To search for pulsations we analyzed data from
the Proportional Counter Array detector \citep[PCA;][]{jsg+96}, acquired
in the {\em GoodXenon\/} mode.  The non-imaging PCA (with a $1\arcdeg$
FWHM field of view) is made up of five counter units that are sensitive to
X-rays in the 2--60\,keV range.  During the observation, photon arrival
times were collected with better than $100\,\mu$s resolution using a
time-weighted average of 3.0 counter units in operation.

\subsection{Parkes Timing Observations}\label{sec:timing}

\psr\ experiences significant ``timing noise'' that biases its position
obtained via timing measurements \citep{cbm+01}.  We have performed
fits to three independent Parkes timing data sets, and while the average
position is consistent with that of \citet{cbm+01}, which we use here,
there is considerable variance, from which we estimate the uncertainty:
(J2000.0) $\mbox{R.A.} = 10^{\rm h}16^{\rm m}21\fs16 \pm 0\fs14$,
$\mbox{Decl.} = -58\arcdeg57'12\farcs1 \pm 0\farcs3$, where the errors
should be considered approximate 68\% confidence level estimates.  It is
possible that future interferometric observations with the Australia
Telescope Compact Array will yield a precise pulsar position free of
significant systematics.

The pulsar has experienced a rotational glitch once since it was
discovered \citep{hfs+04}.  Together with the presence of timing
noise, this required monitoring close to the epoch of the {\em RXTE\/}
observation in order to obtain a good estimate of its period and period
derivative.  Using a data set that included observations obtained two
weeks before and one day after the {\em RXTE\/} observation, we measured a
barycentric $P = 0.1073889238$\,s and $\dot P = 8.077\times10^{-14}$ for
MJD = 53070.210, where the uncertainties are smaller than the precision
given here for these quantities.

\subsection{{\em Chandra\/} Imaging}\label{sec:img}

Figure~\ref{fig:zoom} shows the ACIS-S3 image of a $20''\times20''$
box centered near the position of \psr.  As in all Figures, we restrict
ourselves to the 0.8--7\,keV energy range, since we have found that
outside this range there are very few source counts (\S~\ref{sec:spec}).
In this image we indicate the $1\,\sigma$ error ellipse for the
position of the pulsar determined from radio timing measurements
(\S~\ref{sec:timing}).

\begin{figure*}
\begin{center}
\includegraphics[scale=0.97,angle=270]{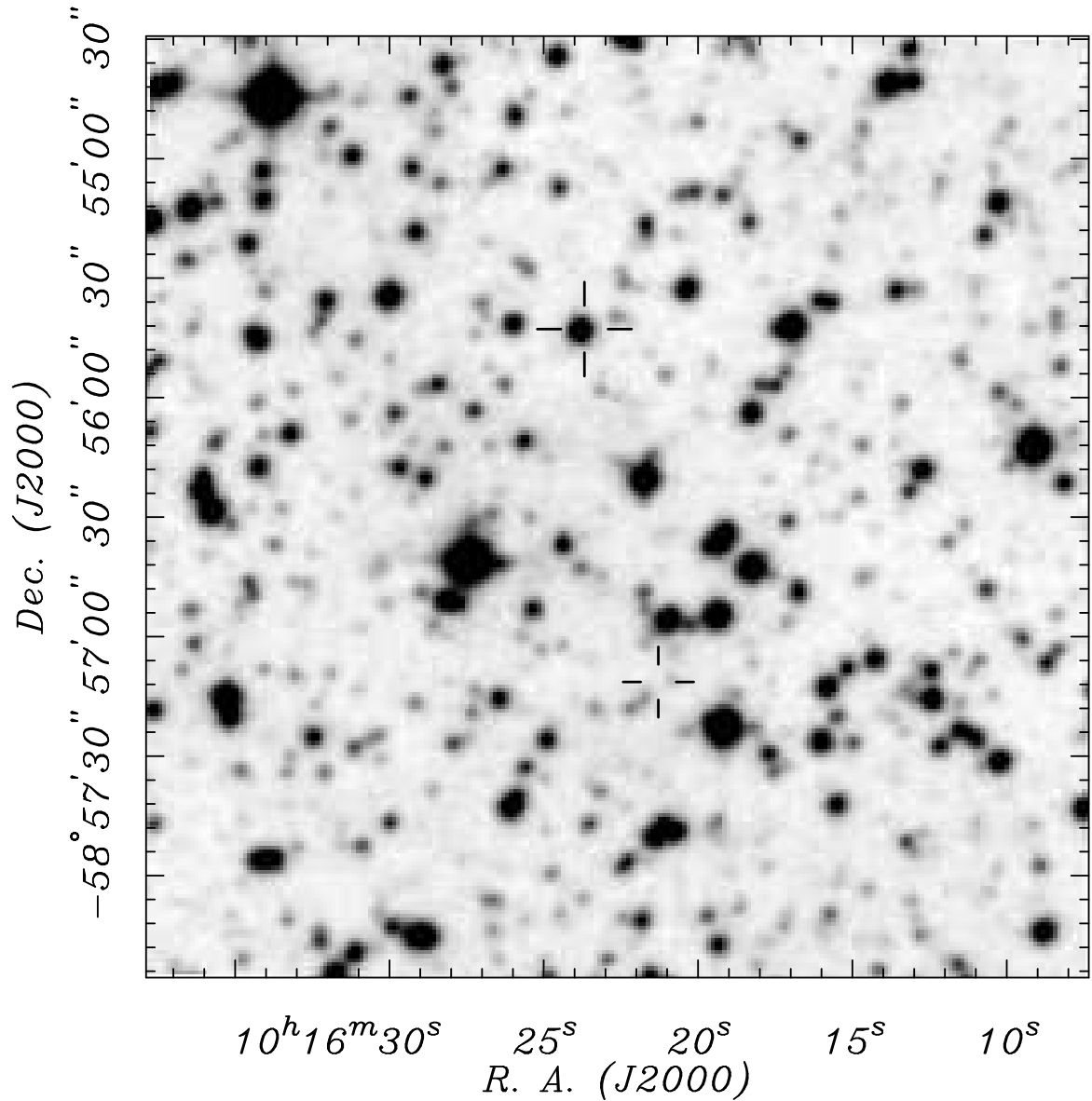}
\caption{ \label{fig:opt_chart} A finding chart from a digitized UK
Schmidt sky survey red plate showing the locations of \psr\ ({\em bottom
cross}) and the serendipitous X-ray source CXOU~J101623.6--585542 ({\em
top cross}). }
\end{center}
\end{figure*}

\begin{figure*}
\begin{center}
\includegraphics[scale=0.64,angle=270]{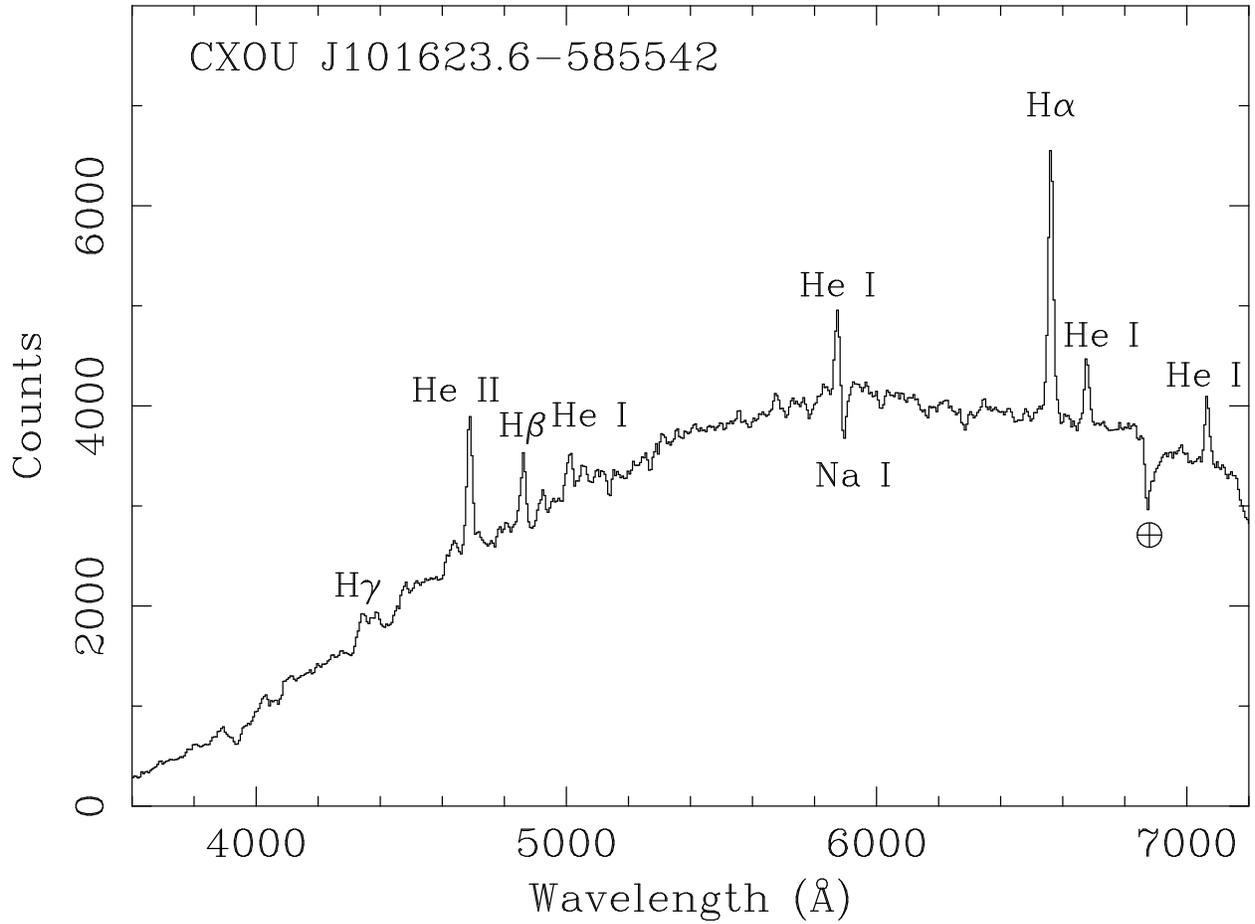}
\caption{ \label{fig:opt_spect} A 30 minute exposure spectrum
of the optical counterpart of the serendipitous X-ray source
CXOU~J101623.6--585542 obtained on the SMARTS/CTIO 1.5\,m telescope on
2004 June 23 (courtesy C. Bailyn).  Flux calibration has not been applied.
Prominent emission lines that are typical of accretion-powered binary
X-ray sources are marked.  There are no obvious stellar photospheric
absorption features.  Interstellar Na~I~D absorption is also seen, as
well as weaker diffuse interstellar bands.  (See Fig.~\ref{fig:opt_chart}
for a finding chart.) }
\end{center}
\end{figure*}

We have checked the {\em Chandra\/} astrometry by comparing the positions
of three X-ray point sources with those of their optical counterparts.  In
all cases there is an excellent match between positions.  In particular,
the optical counterpart of CXOU~J101623.6--585542, the brightest X-ray
point source in the S3 chip (with 300 photons, a hard absorbed spectrum,
and flicker-like variability within the 5\,hr observation), was selected
by positional coincidence from the digitized UK Schmidt plates and the
2MASS survey, at coordinates (2MASS, J2000.0) $\mbox{R.A.} = 10^{\rm
h}16^{\rm m}23\fs66$, $\mbox{Decl.} = -58\arcdeg55'42\farcs7$.  Its
magnitudes from the USNO B1.0 and 2MASS catalogs are $B=15.33, R=14.35,
I=13.94, J=12.95, H=12.51, K=12.35$.  Figure~\ref{fig:opt_chart}, from
the digitized $R$ plate, is a finding chart for this star as well as for
the location of the pulsar.  A low-resolution ($\sim 15\,$\AA) spectrum
of the star was obtained on 2004 June 23 using the RC Spectrograph
and Loral 1K CCD on the SMARTS\footnote{Small and Moderate Aperture
Research Telescope System; see http://www.astro.yale.edu/smarts.}
1.5\,m telescope at the Cerro Tololo Interamerican Observatory.
Three 10 minute exposures were averaged into the raw counts spectrum
shown in Figure~\ref{fig:opt_spect}.  Emission lines of hydrogen and
helium identify this star as an accreting binary, probably a cataclysmic
variable, although possibly a low-mass X-ray binary in quiescence.
The fact that He~II$~\lambda4686$ is stronger than H$\beta$ indicates
high excitation due to X-ray photoionization.  The agreement between
the X-ray and optical positions of this bright source indicates that the
accuracy of the {\em Chandra\/} aspect solution is better than $0\farcs2$
in each coordinate and requires no further refinement.

In Figure~\ref{fig:zoom} we see that the centroid of relatively bright
X-ray emission is not consistent with the nominal pulsar position.
However, considering the systematics involved in obtaining this position,
and at the $3\,\sigma$ level (\S~\ref{sec:timing}), the position of
\psr\ coincides with significant X-ray emission, and is even marginally
consistent with the position of the brightest X-ray pixel (with 14
photons), at (J2000.0) $\mbox{R.A.} = 10^{\rm h}16^{\rm m}21\fs35$,
$\mbox{Decl.} = -58\arcdeg57'11\farcs3$.

Figure~\ref{fig:radial} shows the radial profile of counts extracted
from a circular region $30''$ in radius centered within one pixel of the
brightest X-ray pixel in Figure~\ref{fig:zoom}.  We compared the radial
profile to that expected from a point source as determined by a {\tt
ChaRT} + {\tt MARX} raytrace simulation of the PSF at the appropriate
position on the focal plane\footnote{Following the method outlined in
http://cxc.harvard.edu/chart.}.  The centrally peaked region of emission
is found to be inconsistent with a point source, but rather shows some
extended emission within $2''$ of the central pixel.  This does {\em
not\/} necessarily imply that the pulsar is surrounded by an extended
structure $1''$--$2''$ in size of approximate circular symmetry, since
we do not know the exact pulsar position.  In fact we cannot obviously
decompose this centrally peaked emission into a clear point source
superimposed on a smooth background, and it is possible that the pulsar
is a faint source not resolved against the nebular background.

In Figure~\ref{fig:wide} we show the exposure-corrected image of the
full S3 chip, with smoothing on two different scales (in the same Figure
we also display radio data from the MOST MGPS2 survey\footnote{See
http://www.physics.usyd.edu.au/astrop/most/mgps.html.}, that we discuss
in \S~\ref{sec:disc}).  While X-ray emission drops off rather abruptly to
the SW of the pulsar position, there is a faint ``tongue'' of emission
about $1'\times2'$ in size toward the NE.  This is quantified in
Figure~\ref{fig:boxes}, where we use the standard-processed event files
to extract counts from a rectangular region covering approximately the
tongue, but excluding the bright point source CXOU~J101623.6--585542
to the north of the pulsar, and including a substantial area to
its SW.  Here we see that X-ray emission is brightest in an area of
half-width $\approx 25''$ (that is approximately circular in shape;
cf. Fig.~\ref{fig:wide}), within which it is centrally peaked (see also
Fig.~\ref{fig:radial}).  Also, X-ray emission drops abruptly to the
SW of the pulsar position (negative offsets in the Figure), to a level
that we define as a baseline here, but it remains substantially above
this background toward the NE, for at least $2'$.  Beyond this offset,
mirror vignetting (not corrected for) becomes significant.  In any case,
it is clear from the asymmetry in the profile shown that the tongue
noted in the smoothed Figure~\ref{fig:wide} is a real feature.

\subsection{{\em Chandra\/} Spectroscopy}\label{sec:spec}

As seen in Figure~\ref{fig:boxes}, the majority of the counts near the
pulsar position lie within a region of $\sim 50''$ in extent, that has
approximate circular symmetry (see also Fig.~\ref{fig:wide}).  In order
to obtain a spectrum of this ``core'' region, we extracted counts from
a circular aperture of radius 50 pixels ($r = 25''$) centered around
the pixel with the most photons.  For the background we used an annulus
with the same center, and inner and outer radii of 55 and 150 pixels,
respectively.  Virtually all of the photons in the source region are
in the 0.8--7.0\,keV range, which we used for spectral fits.  In this
energy range there are 367 background-subtracted counts within the $r =
25''$ aperture.

We extracted a spectrum by grouping the 367 counts with at least 30
counts per bin.  We then used the {\sc sherpa} spectral fitting package
to obtain model parameters.  The spectrum is well fitted ($\chi^2_\nu =
0.97$) by a power-law model with photon index $\Gamma = 1.32 \pm 0.25$,
absorbing neutral hydrogen column density $N_{\rm H} = (0.50 \pm 0.17)
\times10^{22}$\,cm$^{-2}$, and unabsorbed flux $F(0.8-7\,{\rm keV}) =
2.8 \times 10^{-13}$\,ergs\,cm$^{-2}$\,s$^{-1}$, where uncertainties are
given at the 68\% confidence level.  Other models (such as black-body)
give fits that are either unphysical (e.g., with an extremely high
temperature of $\sim 1$\,keV and essentially zero $N_{\rm H}$) or
statistically inadequate.

We also extracted counts from a circle with only 10\% the area of the
above extraction region, with $r = 15$ pixels (cf. Fig.~\ref{fig:zoom}).
There are 225 background-subtracted counts in this region, and the best
power-law fit has the same $\Gamma$ and $N_{\rm H}$ as the one made to
the larger area, although because of inadequate statistics the formal
fit is poor.

Finally, we placed an upper limit on emission from the pulsar itself.
We did this by extracting counts from a circle three pixels in
radius centered at various locations near the peak of emission (see
Figs.~\ref{fig:zoom} and \ref{fig:radial}) and finding which circle
contained the greatest flux.  This leads to a very conservative limit,
both because we have no indication that this corresponds to the actual
pulsar position, and also because it assumes no superimposed nebular
emission at the pulsar location, which is unreasonable.  The 114 photons
extracted from this small circular region correspond, assuming the same
spectral parameters as above (absorbed power law with $\Gamma = 1.32$ and
$N_{\rm H} = 5 \times10^{21}$\,cm$^{-2}$), to an unabsorbed pulsar flux
$F(0.8-7\,{\rm keV}) < 9 \times 10^{-14}$\,ergs\,cm$^{-2}$\,s$^{-1}$.


The tongue of emission toward the NE of the pulsar position
(cf. Figs.~\ref{fig:wide} and \ref{fig:boxes}) is too diffuse and
faint for a useful spectral fit to be made, with only $\sim 200$
background-subtracted counts (with a very large uncertainty) in an area
$\sim 2$\,arcmin$^2$.

\subsection{{\em RXTE\/} Search for Pulsations}\label{sec:xte}

The data were processed using the recommended methods and with the
photon arrival times corrected to the solar system barycenter using
the JPL DE200 ephemeris \citep{sta90} for the radio source position
(\S~\ref{sec:timing}).  Selecting good data time intervals using the
standard criteria produced a total of 55.5\,ks of usable data.  In our
PCA timing analysis we selected events from PI channels 2--50 ($\sim
2$--21\,keV) and PCA layer~1 only, optimal for a typical pulsar search,
obtaining a total of $1.27 \times 10^6$ events.

Event times were folded into 20 phase bins at and around the pulsar period
determined from the radio ephemeris (\S~\ref{sec:timing}), and tested
against a flat distribution using a $\chi^2$ statistic. No significant
signal was found in a 1024 bin periodogram within $\pm 2.4\,\mu$s of
the expected period.

\section{Discussion}\label{sec:disc}

\begin{figure*}
\begin{center}
\includegraphics[scale=0.78,angle=270]{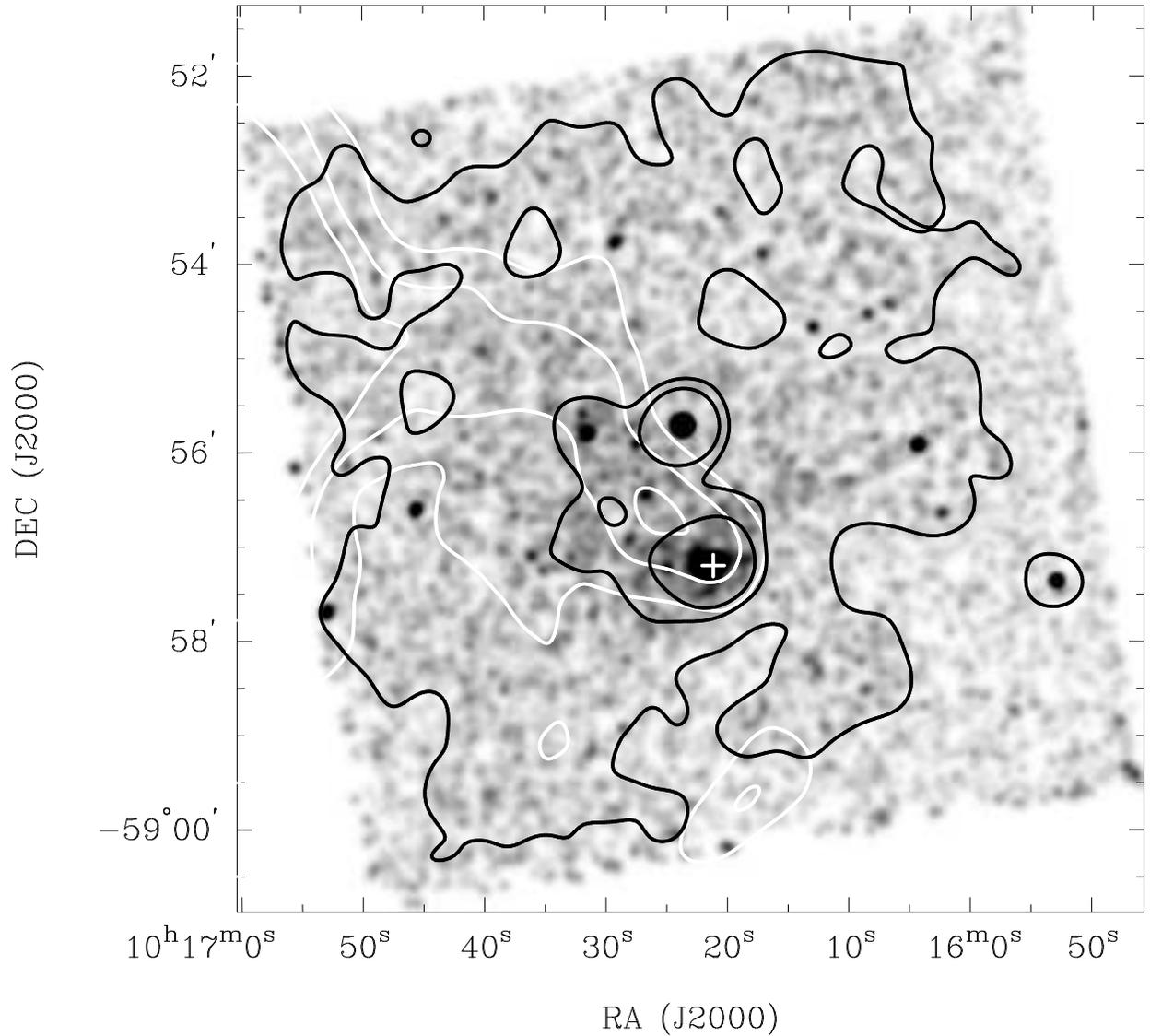}
\caption{ \label{fig:wide} Exposure-corrected {\em Chandra\/} image
in the 0.8--7\,keV range and Molonglo Observatory Synthesis Telescope
(MOST) map at a frequency of 843\,MHz.  Greyscale is X-ray data smoothed
with a Gaussian of FWHM = $7''$, over a linear range of 0\% to 4.4\%
of peak intensity (which occurs in the source CXOU~J101623.6--585542
$\approx 1\farcm5$ to the north of the pulsar).  Black contours are the
same data smoothed with a FWHM = $30''$ Gaussian, at levels of 10\%,
20\%, and 30\% of peak.  White contours are radio data, at levels of 10,
20, and 30 mJy\,beam$^{-1}$, and have a resolution of $43''\times50''$.
A portion of the western edge of the radio SNR~\snr\ is visible toward
the NE corner of the field.  The white cross marks the position of \psr. }
\end{center}
\end{figure*}

\citet{cbm+01} argued that \psr\ is located near an {\em Einstein\/}
X-ray source that could be a PWN.  Also, based on the location of the
pulsar at the tip of a finger of radio emission apparently joined to
the SNR~\snr\ (see Fig.~\ref{fig:wide}), they speculated that the two
objects might be associated.  In that case, the pulsar would be located
at the SNR distance, $d = 3 \pm 0.6$\,kpc \citep{rm86}.  The dispersion
measure of the pulsar, $\mbox{DM} = 394$\,cm$^{-3}$\,pc \citep{cbm+01},
and models for the free electron density, suggest a much larger distance
$d = 9^{+3}_{-2}$\,kpc \citep{tc93} or $d \approx 8$\,kpc \citep{cl02},
although these models can be in error for individual objects by factors
of a few.

\begin{figure*}
\begin{center}
\includegraphics[scale=0.75,angle=270]{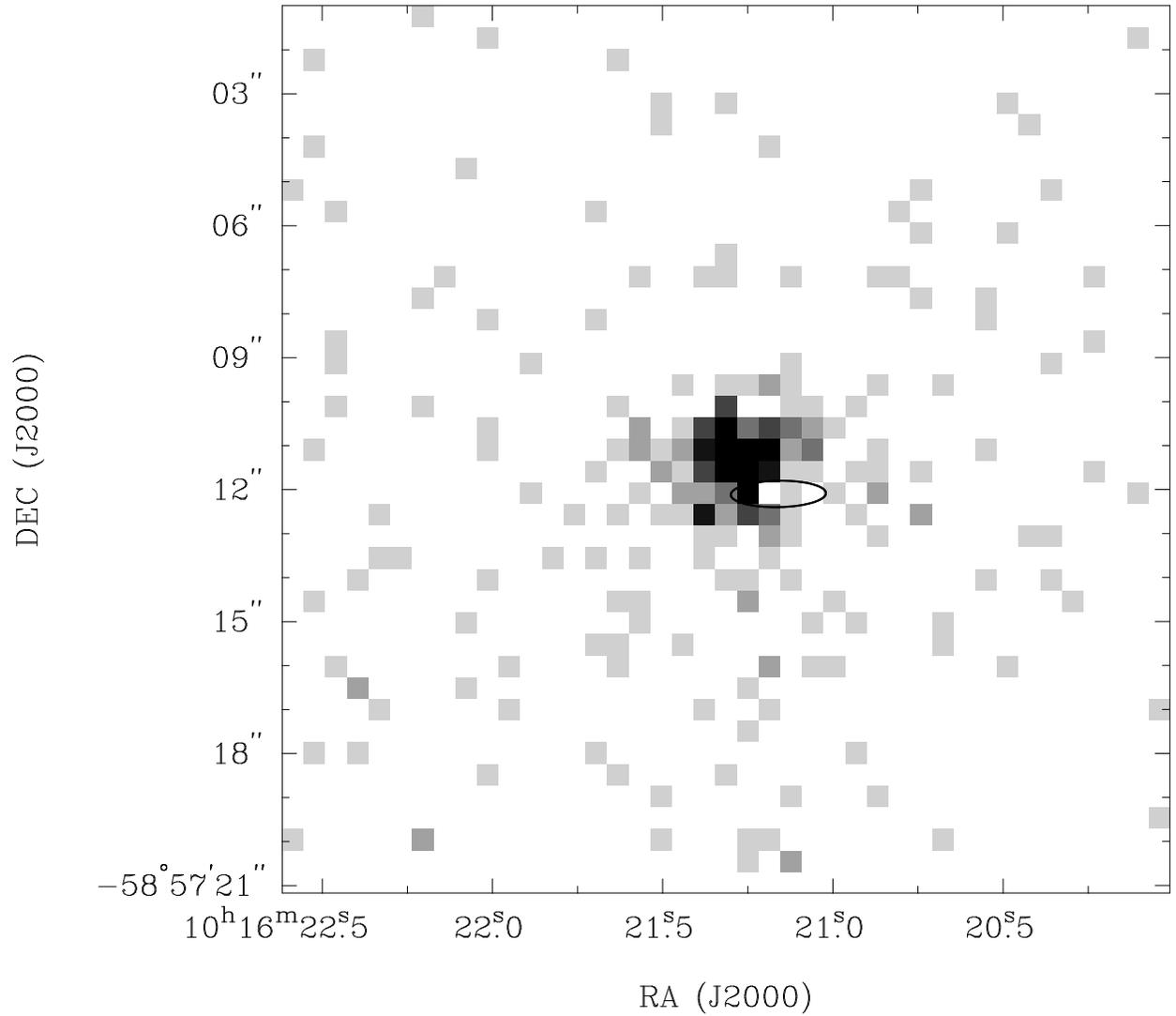}
\caption{ \label{fig:zoom} Exposure-corrected image in the 0.8--7\,keV
range with no smoothing, in $20''\times20''$ box around \psr. Greyscale
is linear, running from 0\% to 50\% of peak in this field.  The ellipse
corresponds to the radio pulsar position and $1\,\sigma$ uncertainties
obtained from timing measurements (\S~\ref{sec:timing}). }
\end{center}
\end{figure*}

The {\em Chandra\/} observation of \psr\ shows that the pulsar is
positionally coincident with a centrally peaked compact X-ray source
(\S~\ref{sec:img}) whose spectrum is well modeled by an absorbed power
law with photon index $\Gamma = 1.3 \pm 0.3$ (\S~\ref{sec:spec}).
Both of these characteristics confirm that \psr\ is associated with a
newly identified PWN that we designate \pwn\ based on its coordinates.
Whether or not the pulsar is exactly coincident with the peak of
X-ray emission (see Fig.~\ref{fig:zoom} and \S~\ref{sec:img}) does
not alter this conclusion \citep[see, e.g.,][]{gpg01}.  Likewise,
while the non-thermal compact source has a harder spectrum than
many PWNe \citep[e.g.,][]{got03}, other well-established PWNe
\citep[e.g.,][]{hgc+02} have photon indices that are consistent, within
the uncertainties, with those of \pwn.  \psr/PWN~\pwn\ joins about
13 Vela-like pulsars now detected at X-ray energies \citep[see][and
references therein]{gsk+03,pccm02}.

The apparent X-ray source identified in {\em Einstein\/} data \citep[where
a $3'$ extraction radius was used;][]{cbm+01}, was about 3 times brighter
than what we infer for PWN~\pwn\ (including the faint diffuse emission).
Part of this discrepancy arises from the inclusion of the bright point
source CXOU~J101623.6--585542 (Fig.~\ref{fig:wide}) in the {\em Einstein\/}
extraction region, but after accounting for this a factor of $\sim 2$
remains unexplained.  This point source is likely an accreting binary
(see \S~\ref{sec:img}), unrelated to the PWN, and is variable within
the {\em Chandra\/} observation, providing a possible explanation for
the remaining discrepancy.

The neutral hydrogen column density obtained from the
spectral fits (\S~\ref{sec:spec}), $N_{\rm H} = (0.50 \pm 0.17)
\times10^{22}$\,cm$^{-2}$, is one-third of the total Galactic value
in this direction estimated from the H{\sc i} study of \citet{dl90},
$N_{\rm H} \approx 1.50 \times10^{22}$\,cm$^{-2}$.  This argues that \psr\
is located substantially nearer than the DM-derived $d \sim 9$\,kpc:
at Galactic latitude $b = -1\fdg88$, the height above the plane would
be 300\,pc, essentially above the Galactic dust layer, leading one to
expect the maximum neutral column density.  While the conversion of column
density to distance is crude, and our $N_{\rm H}$ value has a significant
uncertainty, the X-ray data are consistent with $d \sim 3$\,kpc, and
hereafter we parametrize distance by $d_3 = d/(3\,\mbox{kpc})$.

In \S~\ref{sec:spec} we obtained X-ray fluxes in the energy range of the
detected photons, 0.8--7\,keV.  Hereafter we quote spectral quantities in
the 2--10\,keV range.  The isotropic X-ray luminosities obtained from our
data are $L_{\rm X\,pwn} = 3 \times 10^{32}\,d_3^2$\,ergs\,s$^{-1}$,
and $L_{\rm X\,psr} < 9 \times 10^{31}\,d_3^2$\,ergs\,s$^{-1}$.
In terms of the pulsar spin-down luminosity, these luminosities
are $L_{\rm X\,pwn}/\dot E = 1.1 \times 10^{-4}\,d_3^2$ and $L_{\rm
X\,psr}/\dot E < 3.5 \times 10^{-5}\,d_3^2$.  This efficiency and upper
limit for conversion of rotational kinetic energy to X-rays are lower
than for many PWNe/pulsars \citep[e.g.,][]{pccm02}, but are similar
to those of Vela \citep{hgh01,pzs+01} and other Vela-like objects
\citep[e.g.,][]{hcg+01}.  The non-detection of pulsations from \psr\ in
the {\em RXTE\/} observation (\S~\ref{sec:xte}) is consistent with this
upper limit on unpulsed emission:  assuming a best-case scenario with
pulsed fraction of 100\% and duty cycle of 5\%, we would have expected
an {\em RXTE\/} signal-to-noise ratio $\la 6$.

The EGRET source \egr\ has a photon index $\Gamma = 2.32 \pm 0.13$
\citep{hbb+99} that is consistent with those of some Vela-like pulsars
\citep[e.g.,][]{hcg+01}, apparently has a steady flux \citep{mmct96},
as is the case for other pulsar-powered $\gamma$-ray sources, and
its isotropic luminosity above 100\,MeV is $L_{\gamma}/\dot E =
0.06\,d_3^2$ \citep{cbm+01}.  Using the {\em Chandra\/} observation,
$L_{\gamma}/L_X \sim 500$, and \psr\ remains a viable candidate for
generating the $\gamma$-rays detected from \egr.  Pulsars are the only
established class of Galactic objects that power EGRET sources, but
CXOU~J101623.6--585542 (see \S~\ref{sec:img}), a relatively bright and
hard X-ray source whose exact nature is unknown, should probably not
be dismissed out of hand as a possible contributor to \egr.  These two
are the brightest X-ray sources in the {\em Einstein\/} field of this
region \citep[cf.][]{cbm+01}.  However, that observation covered only
about 1/3 of the area of the \egr\ error box (the {\em Chandra\/} S3
chip covers only 1\% of the box, and the roll angle of the observation
was such that most other ACIS CCDs did not fall within the EGRET box),
and it is possible that another heretofore undetected X-ray source is
in fact responsible for \egr.  An incontrovertible conclusion on this
matter will likely require the {\em GLAST\/} observatory.

\begin{figure*}
\begin{center}
\includegraphics[scale=0.68,angle=270]{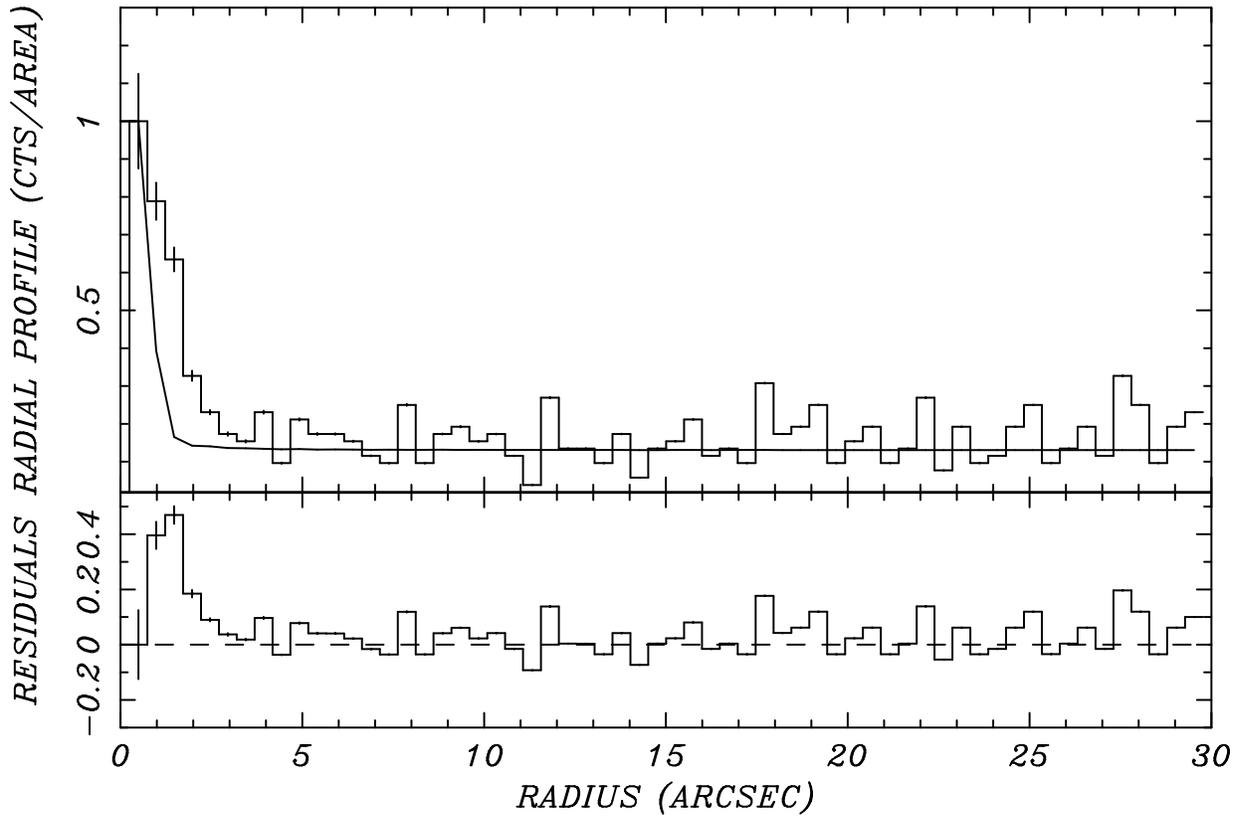}
\caption{ \label{fig:radial} {\em Top\/}: Radial profile of 0.8--7\,keV
counts centered approximately on the peak of emission. The solid line
corresponds to the computed PSF, arbitrarily normalized to the central
bin. {\em Bottom\/}: Residuals of data minus PSF, showing extended
emission within $\sim 2''$ of the center. }
\end{center}
\end{figure*}

As noted in \S~\ref{sec:img}, the brightest component of PWN \pwn\
(Fig.~\ref{fig:zoom}), superimposed on fainter emission, is not
consistent with a point source (Fig.~\ref{fig:radial}).  By analogy
with the structures now observed surrounding many young and energetic
pulsars \citep[see, e.g.,][]{nr04}, we {\em speculate\/} that this
structure may indicate the pulsar wind termination shock.  If so,
$r_s \approx 2'' = 0.03\,d_3$\,pc, similar in size to that of the Vela
pulsar \citep{hgh01}.  Proceeding further with this interpretation
one runs into familiar problems: a simple estimation of the nebular
magnetic field downstream of the shock obtained from the assumption
of equipartition suggests a few tens of $\mu$G, and the corresponding
synchrotron lifetime of electrons/positrons at $\sim 1.5$\,keV (the peak
of our observed energy distribution) is so short (a few hundred years or
less) that the very high velocity required to transport the particles
to the edge of the PWN is at variance with more reasonable nebular
expansion speeds \citep[e.g.,][]{kc84,vagt01}.  In any case, despite
the limitations of existing X-ray data on \psr/PWN~\pwn\ that prevent
further quantitative analysis, the system shows several characteristics
similar to, or consistent with, those of other Vela-like pulsars/PWNe.

\begin{figure*}
\begin{center}
\includegraphics[scale=0.82,angle=0]{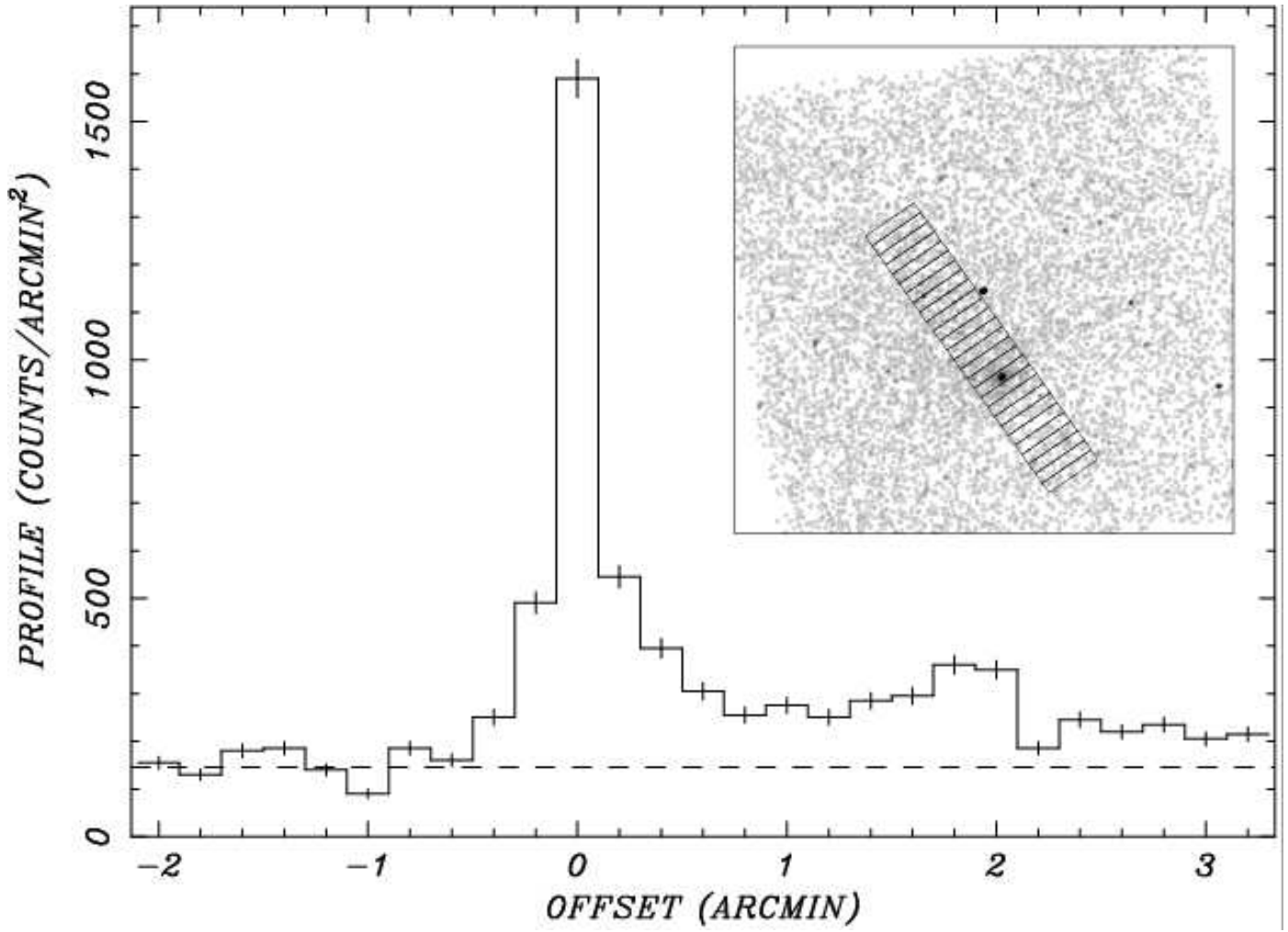}
\caption{ \label{fig:boxes} Counts extracted from 27 adjacent rectangular
boxes (see inset, with north to the top and east to the left), each
$1'$ in width and $0\farcm2$ in length, with the long axis of the
overall $1'\times 5\farcm4$ area oriented at $35\arcdeg$ east of north.
Zero offset corresponds to the box centered on the position of \psr,
with negative offsets in the SW direction and positive offsets in
the NE direction.  The slight increase in counts seen in the two bins
near an offset of $2'$ is contributed by a point source. Compare to
Figure~\ref{fig:wide}.  }
\end{center}
\end{figure*}

We now discuss the tongue of X-ray emission that has \psr\ and its
$\sim 25''$ PWN located near its SW end (Figs.~\ref{fig:wide} and
\ref{fig:boxes}), as well as the possible association of the pulsar
with SNR~\snr.  While the $\approx 1'\times2'$ X-ray feature is real
(\S~\ref{sec:img}), we have no a priori expectation that it should be
associated with the pulsar.  The pulsar and this extended feature are both
located, in projection, $\sim 4'$ to the west of the bright (in radio)
western edge of SNR~\snr\ (see Fig.~\ref{fig:wide}).  The SNR in turn
is interacting with a molecular cloud \citep{rm86}.  It is therefore
possible that the very faint and extended X-ray emission is somehow
related to the SNR/molecular cloud and not to the pulsar.

On the other hand, as mentioned before, the pulsar is
located at the SW tip of a relatively bright finger of radio
emission that appears to connect at its NE end to the SNR
\citep[Fig.~\ref{fig:wide};][]{mck+89,cbm+01}.  It is notable that the
X-ray tongue provides a good match to the SW tip of the radio finger,
in both width and position angle --- with the pulsar also located very
near the SW end of X-ray emission.  These coincidences are suggestive
of a possible connection between the radio finger and X-ray tongue,
and in turn of the pulsar with both of them.  The length of the
tongue NE of the pulsar is $L \approx 2' = 1.8\,d_3$\,pc, with a SW
extension $\le 0\farcm2$ (Fig.~\ref{fig:boxes}).  If the X-ray tongue is
related to the pulsar, its cause is not clear, since it is not easy to
understand how such a large distance could be covered within the short
synchrotron lifetimes of the radiating particles.  It has been suggested
\citep[e.g.,][]{wlb93} that wind acceleration can occur to generate very
large velocities, but this should result in visible collimation, which we
do not observe here.  Also, in cases where pulsars travel supersonically
through the ambient medium, a bow shock develops due to ram pressure
with the pulsar at one end of what is often a bright elongated PWN that
gradually fades opposite the direction of motion \citep[e.g.,][]{gvc+04}.
However, in the present case the pulsar velocity required would likely
be several 1000\,km\,s$^{-1}$, and in any case the X-ray emission again
does not appear to be very collimated as would be expected.  If somehow
such a large velocity can explain these observations, then the pulsar
would presumably be traveling due SW, rather than due west as might be
expected if it had originated from the apparent center of SNR~\snr, a
reminder that even if the pulsar is related to the radio finger and X-ray
tongue, the connection to the SNR would not necessarily be established,
since the objects could appear close merely in projection.

The \psr/PWN~\pwn\ system shows significant parallels to
PSR~B1823--13/PWN~G18.0--0.7, another Vela-like pulsar with parameters
very similar to those of J1016--5857.  A deep {\em XMM\/} observation of
B1823--13 \citep{gsk+03} shows that the pulsar and its compact PWN are
embedded asymmetrically within a fainter and much larger nebula whose
origin is not entirely clear (two potentially important differences
between the systems are that no SNR is detected near B1823--13,
and also that its inferred wind-termination shock radius is an order
of magnitude larger than for J1016--5857).  Further observations are
required in order to understand fully the structures observed near \psr\
and their relationships: radio observations may reveal in detail whether
the radio finger is related to the SNR, and whether the finger emission
is non-thermal or polarized, and how it evolves along its length toward
the pulsar at its end. X-ray observations may reveal useful spectral
information and show whether the extended emission, if non-thermal, has
a softer spectrum than that of the compact PWN near the pulsar position
due to synchrotron losses, and help elucidate its origin.

\acknowledgments

We are grateful to Charles Bailyn for obtaining, using the SMARTS 1.5\,m
telescope, the optical data described in \S~\ref{sec:img}.  We thank Vicky
Kaspi for early contributions to this work.  We are grateful to Marta
Burgay and Andy Faulkner for helping with data collection at Parkes.
The Parkes Observatory is part of the Australia Telescope, which is
funded by the Commonwealth of Australia for operation as a National
Facility managed by CSIRO.  We thank Dick Hunstead and Anne Green for
supplying the radio image used in Figure~\ref{fig:wide}.  The Molonglo
Observatory Synthesis Telescope is supported by the Australian Research
Council.  FC thanks Caleb ``Tweety Bird'' Scharf for his irrepressible
good humor and patience.  This work was supported by NASA through SAO
grant GO3-4076X.


\clearpage

\clearpage

\clearpage

\clearpage

\clearpage

\clearpage

\begin{thebibliography}{25}
\expandafter\ifx\csname natexlab\endcsname\relax\def\natexlab#1{#1}\fi

\bibitem[{Camilo {et~al.}(2001)Camilo, Bell, Manchester, Lyne, Possenti,
  Kramer, Kaspi, Stairs, D'Amico, Hobbs, Gotthelf, \& Gaensler}]{cbm+01}
Camilo, F., et al. 2001, ApJ, 557, L51

\bibitem[{Cordes \& Lazio(2002)}]{cl02}
Cordes, J.~M., \& Lazio, T.~J.~W. 2002, preprint (astro-ph/0207156)

\bibitem[{Dickey \& Lockman(1990)}]{dl90}
Dickey, J.~M., \& Lockman, F.~J. 1990, ARA\&A, 28, 215

\bibitem[{Gaensler {et~al.}(2001)Gaensler, Pivovaroff, \& Garmire}]{gpg01}
Gaensler, B.~M., Pivovaroff, M.~J., \& Garmire, G.~P. 2001, ApJ, 556, L107

\bibitem[{{Gaensler} {et~al.}(2003){Gaensler}, {Schulz}, {Kaspi}, {Pivovaroff},
  \& {Becker}}]{gsk+03}
{Gaensler}, B.~M., {Schulz}, N.~S., {Kaspi}, V.~M., {Pivovaroff}, M.~J., \&
  {Becker}, W.~E. 2003, ApJ, 588, 441

\bibitem[{Gaensler {et~al.}(2004)Gaensler, van~der Swaluw, Camilo, Kaspi,
  Baganoff, Yusef-Zadeh, \& Manchester}]{gvc+04}
Gaensler, B.~M., van~der Swaluw, E., Camilo, F., Kaspi, V.~M., Baganoff, F.~K.,
  Yusef-Zadeh, F., \& Manchester, R.~N. 2004, ApJ, in press (astro-ph/0312362)

\bibitem[{{Gotthelf}(2003)}]{got03}
{Gotthelf}, E.~V. 2003, ApJ, 591, 361

\bibitem[{{Halpern} {et~al.}(2001){Halpern}, {Camilo}, {Gotthelf}, {Helfand},
  {Kramer}, {Lyne}, {Leighly}, \& {Eracleous}}]{hcg+01}
{Halpern}, J.~P., {Camilo}, F., {Gotthelf}, E.~V., {Helfand}, D.~J., {Kramer},
  M., {Lyne}, A.~G., {Leighly}, K.~M., \& {Eracleous}, M. 2001, ApJ, 552, L125

\bibitem[{{Halpern} {et~al.}(2002){Halpern}, {Gotthelf}, {Camilo}, {Collins},
  \& {Helfand}}]{hgc+02}
{Halpern}, J.~P., {Gotthelf}, E.~V., {Camilo}, F., {Collins}, B., \& {Helfand},
  D.~J. 2002, in Neutron Stars in Supernova Remnants, eds. P.~O. Slane \&
  B.~M. Gaensler (San Francisco: ASP), 199

\bibitem[{Hartman {et~al.}(1999)Hartman, Bertsch, Bloom, Chen, Deines-Jones,
  Esposito, Fichtel, Friedlander, Hunter, McDonald, Sreekumar, Thompson, Jones,
  Lin, Michelson, Nolan, Tompkins, Kanbach, Mayer-Hasselwander, M\"ucke, Pohl,
  Reimer, Kniffen, Schneid, von Montigny, Mukherjee, \& Dingus}]{hbb+99}
Hartman, R.~C., et al. 1999, ApJS, 123, 79

\bibitem[{{Helfand} {et~al.}(2001){Helfand}, {Gotthelf}, \& {Halpern}}]{hgh01}
{Helfand}, D.~J., {Gotthelf}, E.~V., \& {Halpern}, J.~P. 2001, ApJ, 556, 380

\bibitem[{{Hobbs} {et~al.}(2004){Hobbs}, {Faulkner}, {Stairs}, {Camilo},
  {Manchester}, {Lyne}, {Kramer}, {D'Amico}, {Kaspi}, {Possenti}, {McLaughlin},
  {Lorimer}, {Burgay}, {Joshi}, \& {Crawford}}]{hfs+04}
{Hobbs}, G., et al. 2004, MNRAS, in press (astro-ph/0405364)

\bibitem[{Jahoda {et~al.}(1996)Jahoda, Swank, Giles, Stark, Strohmayer, Zhang,
  \& Morgan}]{jsg+96}
Jahoda, K., Swank, J.~H., Giles, A.~B., Stark, M.~J., Strohmayer, T., Zhang,
  W., \& Morgan, E.~H. 1996, Proc. SPIE, 2808, 59

\bibitem[{Kennel \& Coroniti(1984)}]{kc84}
Kennel, C.~F., \& Coroniti, F.~V. 1984, ApJ, 283, 710

\bibitem[{Manchester {et~al.}(2001)Manchester, Lyne, Camilo, Bell, Kaspi,
  D'Amico, McKay, Crawford, Stairs, Possenti, Morris, \& Sheppard}]{mlc+01}
Manchester, R.~N., et al. 2001, MNRAS, 328, 17

\bibitem[{McLaughlin {et~al.}(1996)McLaughlin, Mattox, Cordes, \&
  Thompson}]{mmct96}
McLaughlin, M.~A., Mattox, J.~R., Cordes, J.~M., \& Thompson, D.~J. 1996, ApJ,
  473, 763

\bibitem[{Milne {et~al.}(1989)Milne, Caswell, Kesteven, Haynes, \&
  Roger}]{mck+89}
Milne, D.~K., Caswell, J.~L., Kesteven, M.~J., Haynes, R.~F., \& Roger, R.~S.
  1989, Proc. Astr. Soc. Aust., 8, 187

\bibitem[{{Ng} \& {Romani}(2004)}]{nr04}
{Ng}, C.-Y., \& {Romani}, R.~W. 2004, ApJ, 601, 479

\bibitem[{Pavlov {et~al.}(2001)Pavlov, Zavlin, Sanwal, Burwitz, \&
  Garmire}]{pzs+01}
Pavlov, G.~G., Zavlin, V.~E., Sanwal, D., Burwitz, V., \& Garmire, G.~P. 2001,
  ApJ, 552, L129

\bibitem[{Possenti {et~al.}(2002)Possenti, Cerutti, Colpi, \&
  Mereghetti}]{pccm02}
Possenti, A., Cerutti, R., Colpi, M., \& Mereghetti, S. 2002, A\&A, 387, 993

\bibitem[{Ruiz \& May(1986)}]{rm86}
Ruiz, M.~T., \& May, J. 1986, ApJ, 309, 667

\bibitem[{Standish(1990)}]{sta90}
Standish, E.~M. 1990, A\&A, 233, 252

\bibitem[{Taylor \& Cordes(1993)}]{tc93}
Taylor, J.~H., \& Cordes, J.~M. 1993, ApJ, 411, 674

\bibitem[{van~der Swaluw {et~al.}(2001)van~der Swaluw, Achterberg, Gallant, \&
  T\'{o}th}]{vagt01}
van~der Swaluw, E., Achterberg, A., Gallant, Y.~A., \& T\'{o}th, G. 2001, A\&A,
  380, 309

\bibitem[{Wang {et~al.}(1993)Wang, Li, \& Begelman}]{wlb93}
Wang, Q.~D., Li, Z.-Y., \& Begelman, M.~C. 1993, Nature, 364, 127

\end{thebibliography}
\end{document}